\documentclass[journal]{IEEEtran}
\usepackage[OT1]{fontenc} 
\usepackage{cite}
\usepackage{float}
\usepackage{multicol,multirow}
\usepackage{makecell,booktabs}
\usepackage{hyperref}
\usepackage[inkscapeformat=pdf]{svg}
\hypersetup{    
    colorlinks=true,
    linkcolor=black,
    anchorcolor=black,
    citecolor=black,
    filecolor=black,
    menucolor=black,
    runcolor=black,
    urlcolor=black,
}

\usepackage{tikz}
\usepackage[utf8]{inputenc}
\usepackage{pgfplots}
\usepgfplotslibrary{groupplots,dateplot}
\usetikzlibrary{patterns,shapes.arrows}
\pgfplotsset{compat=newest}

\usepackage{balance}
\usepackage{xcolor}
\usepackage{colortbl}
\usepackage{graphicx}
\usepackage{amsfonts}
\usepackage{amssymb}
\usepackage{amsmath}
\usepackage{amsthm}
\usepackage{mathtools}
\usepackage{array}
\usepackage{textcomp}
\newcolumntype{P}[1]{>{\centering\arraybackslash}p{#1}}
\newcolumntype{L}[1]{>{\raggedright\arraybackslash}p{#1}}
\newcolumntype{R}[1]{>{\raggedleft\arraybackslash}p{#1}}
\usepackage{color}
\usepackage{epstopdf}
\usepackage{bbm}
\usepackage{bm}
\usepackage{comment}
\usepackage{forest}
\usepackage{nomencl}
\usepackage{enumitem,kantlipsum}
\newtheorem{theorem}{Theorem}
\usepackage{etoolbox}
\usepackage{subcaption}
\usepackage{booktabs}
\usepackage{siunitx}
\usepackage[ruled]{algorithm2e}

\usepackage{euscript}

\newtheorem{proposition}[theorem]{Proposition}

\newtheorem{corollary}[theorem]{Corollary}
\newtheorem{lemma}[theorem]{Lemma}

\usepackage{alphalph,etoolbox} 
\patchcmd{\subequations}{\alph{equation}}{\alphalph{\value{equation}}}{}{}

\allowdisplaybreaks

\begin{document}
\title{Integrating Conductor Health into Dynamic Line Rating and Unit Commitment under Wind Uncertainty}
\author{
    Geon~Roh,~\IEEEmembership{Student Member,~IEEE,}
    and Jip~Kim,~\IEEEmembership{Member,~IEEE}
    \thanks{    
    G. Roh and J. Kim are with Department of Energy Engineering, KENTECH, South Jeolla, 58830 Republic of Korea.
    } 
}

\maketitle
\begin{abstract}
    Dynamic line rating (DLR) enables greater utilization of existing transmission lines by leveraging real-time weather data. 
    However, the elevated temperature operation (ETO) of conductors under DLR, particularly in the presence of uncertainty, is often overlooked, despite its long-term impact on conductor health. 
    This paper addresses ETO under DLR and wind power uncertainty by 1) quantifying risk-based depreciation costs associated with ETO, 2) characterizing correlation-driven ETO risk from wind power and DLR forecast errors, and 3) proposing a Conductor Health-Aware Unit Commitment (CHA-UC) that internalizes these costs in operational decisions. 
    CHA-UC incorporates a robust linear approximation of conductor temperature and integrates expected depreciation costs due to hourly ETO into the objective function. 
    Case studies on the Texas 123-bus backbone test system demonstrate that the proposed CHA-UC model reduces the total cost by 0.75\% and renewable curtailment by 82\% compared to static line rating (SLR) and outperforms quantile regression forest-based methods, while conventional DLR operation without risk consideration resulted in higher costs due to excessive ETO. 
    Further analysis shows that CHA-UC achieves safer line utilization by shifting generator commitments and endogenously adapting to uncertainty correlation, relaxing flows under risk-hedging conditions and tightening flows under risk-amplifying conditions.
\end{abstract}

\section*{Nomenclature}
\subsection*{Sets and Indices}
\addcontentsline{toc}{subsection}{Sets and Indices}
\begin{IEEEdescription}[\IEEEusemathlabelsep\IEEEsetlabelwidth{$s(l),\,r(l)$}]
  \item[$b\in\mathcal{B}$] Set of buses.
  \item[$l\in \mathcal{L}$]{Set of all transmission lines where $\mathcal{L}^\text{DLR/SLR}$ denote sets of DLR/SLR applied lines
  }
  \item[$i\in\mathcal{I}$] Set of thermal generating units.
  \item[$o\in\mathcal{O}_i$] Set of startup segments for unit~$i$.
  \item[$w\in \mathcal{W}_b$] Set of wind farms connected at bus~$b$.
  \item[$t\in\mathcal{T}$] Set of dispatch time periods.
  \item[$s\in\mathcal{S}$] Set of uncertainty scenarios.
  \item[$s(l),\,r(l)$] Sending‐ and receiving‐end buses of line~$l$.
\end{IEEEdescription}

\subsection*{Parameters}
\addcontentsline{toc}{subsection}{Parameters}
\begin{IEEEdescription}[\IEEEusemathlabelsep\IEEEsetlabelwidth{$T^\mathrm{SU}_{i,o},\,T_{i,o+1}^\mathrm{SU}$}]
  \item[$A_{l,t}$, $B_{l,t}$] Slope and intercept of linear fit of loading-temperature for line $l$ [°C/MW, °C].
  \item[$\hat{A}_{l,t}$, $\hat{B}_{l,t}$] DA forecast of slope and intercept of the linear fit for line $l$ at time $t$ [°C/MW, °C].
  \item[$C^{\rm Dep}_{l,t}(\cdot)$] Line depreciation cost function [\$/°C].
  \item[$C^{\mathrm P}_i(\cdot)$] Quadratic fuel cost function [\$/MWh]. 
  \item[$C^{\mathrm{r}+}_i,\,C^{\mathrm{r}-}_i$] Up and down‐reserve costs [\$/MWh].
  \item[$C^{\mathrm{SU}}_{i,o},\,C^{\mathrm{SD}}_i$] Startup and shutdown costs [\$/MWh].
  \item[$C^{\rm VoLL}$] Value of lost load [\$/MWh].
  \item[$\underline{F}^{\mathrm{SLR}}_l,\,\overline{F}^{\mathrm{SLR}}_l$] SLR of line~$l$ [MW].
  \item[$\underline{F}^{\mathrm{DLR}}_{l,t},\,\overline{F}^{\mathrm{DLR}}_{l,t}$] RT DLR of line~$l$, time~$t$ [MW].
  \item[$\hat{\underline{F}}^{\mathrm{DLR}}_{l,t},\,\hat{\overline{F}}^{\mathrm{DLR}}_{l,t}$] DA DLR forecasts of line~$l$, time~$t$ [MW].
  \item[$\underline{G}_i,\,\overline{G}_i$] Minimum, maximum power output [MW].
  \item[$P^{\rm d}_{b,t}$] Load at bus~$b$ and time~$t$ [MW].
  \item[$P^{\rm wind}_{w,t}$] RT WP generation [MW].
  \item[$\hat{P}^{\rm wind}_{w,t}$] DA Forecast WP generation [MW].
  \item[$\Pr_s$] Probability of scenario~$s$.
  \item[$R^{\text{min},+}_t$] Required system total up reserve.
  \item[$RU_i,\,RD_i$] Ramp‐up and ramp‐down limits [MW/h].
  \item[$SU_i,\,SD_i$] Startup and shutdown capability [MW].
  \item[$T^\mathrm{SU}_{i,o}$] Hours defining the segment $o$ limits [h].
  \item[$TU_i,\,TD_i$] Minimum up‐ and down‐times of unit~$i$ [h].
  \item[$X_l$] Reactance of line~$l$ [p.u.].
  \item[$\underline{\Theta}_l,\,\overline{\Theta}_l$] Phase‐angle difference limits on line~$l$ [rad].
  \item[$\xi_{l,t,s}$] DLR forecast error scenario $s$ [MW].
  \item[$\omega_{w,t,s}$] WP forecast error scenario $s$ [MW].
  \item[$K_{l,t}$] DLR security margin multiplier.
\end{IEEEdescription}

\subsection*{Variables}
\addcontentsline{toc}{subsection}{Variables}
\begin{IEEEdescription}[\IEEEusemathlabelsep\IEEEsetlabelwidth{$\hat{p}^{\rm cur,adj}_{w,t,s}$,\,$p^{\rm cur,adj}_{w,t}$}]
  \item[$g_{i,t}$] DA dispatch of unit~$i$ at time~$t$ [MW].
  \item[$\hat{g}^{\rm adj}_{i,t,s}$, \, $g^{\rm adj}_{i,t}$] DA and RT adjusted generation of unit~$i$ at time~$t$ [MW].
  \item[$p_{l,t}$] DA power flow on line~$l$ at time~$t$ [MW].
  \item[$\hat{p}^{\rm adj}_{l,t,s}$,\,$p^{\rm adj}_{l,t}$] DA and RT adjusted power flow [MW].
  \item[$p^{\rm cur}_{w,t}$] DA wind generation curtailment [MW].
  \item[$\hat{p}^{\rm cur,adj}_{w,t,s}$, $p^{\rm cur,adj}_{w,t}$] DA and RT adjusted wind generation curtailment at farm~$w$, time~$t$ [MW].
  \item[$\hat{p}^{\rm shed}_{b,t}$] DA load shedding at bus~$b$, time~$t$ [MW].
  \item[$\hat{p}^{\rm shed, adj}_{b,t,s}$, $p^{\rm shed, adj}_{b,t}$] DA, RT adjusted load shedding [MW].
  \item[$r^{\text{avail},+}_{i,t}$] DA available online up reserve [MW].
  \item[$\hat{r}^+_{i,t,s},\hat{r}^-_{i,t,s}$] DA up and down‐reserve activation given scenario~$s$ for unit~$i$, time~$t$, [MW].
  \item[$r^+_{i,t},r^-_{i,t}$] RT up, down‐reserve activation [MW].
  \item[$u_{i,t}$] Binary commitment variable
  \item[$u_{i,t}^{\rm SU}$,~$u_{i,t}^{\rm SD}$] Binary startup/shutdown variable. 
  \item[$\delta_{i,o,t}$] Activation of startup‐segment $o$.
  \item[$\theta_{b,t}$] DA voltage angle [rad].
  \item[$\hat{\theta}^{\rm adj}_{b,t,s}$,$\theta^{\rm adj}_{b,t}$] DA and RT adjusted voltage angle [rad].
  \item[$\hat{\tau}_{l,t,s}$] DA conductor temperature given scenario $s$ on line~$l$, time~$t$ [°C].
  \item[$\tau_{l,t}$] RT temperature of conductor [°C].
\end{IEEEdescription}

\section{Introduction}\label{Sec:Intro}

The large-scale integration of renewable energy sources (RES) into power systems remains a major technical challenge, particularly due to the spatial mismatch between high-potential generation sites and demand centers.
Areas with high RES potential are far from demand, and efficient RES integration plans entail intense transmission expansion \cite{Impram2020}. 
Case studies in Texas indicate that transmission expansion costs rise significantly when RES are integrated due to their remote locations \cite{Herding2024}.
Moreover, transmission construction lags generation construction: wind and solar projects can be built within three years, whereas new transmission requires five to ten years \cite{Donoho2015, Clapin2024}.
This mismatch causes increased operation costs, renewable generation curtailment, and grid congestion.
Given these challenges, maximizing the use of existing transmission lines is essential for RES integration.

Dynamic line rating (DLR) has emerged as a promising solution that exploits real-time weather conditions to increase the utilization of existing transmission facilities. 
Unlike static line rating (SLR), which assumes conservative worst-case weather conditions (e.g., 0.5 m/s wind speed, 40°C ambient temperature), DLR adjusts ampacity dynamically. The synergy with wind power is particularly strong, as both depend on wind speed, a spatially correlated factor \cite{Glaum2023}.
Authors of \cite{Wallnerstrom2015} and \cite{Lee2022} both prove that applying DLR is effective in reducing WP curtailment. 
Especially, \cite{Lee2022} shows that the system operator may save 776 M\$ per year from congestion costs in Texas by introducing DLR. 
In 2024, FERC issued an Advance Notice of Proposed Rulemaking on DLR \cite{FERC2024}, building on Order 881 \cite{FERC2021}, which expanded line ratings beyond ambient-adjusted approaches. These actions foreshadow the expanding implementation and importance of DLR.

However, DLR can increase the risk of accelerated conductor degradation, particularly under elevated temperature operation (ETO) and in corrosive environments. 
Thermal annealing and corrosion are identified as key factors limiting conductor life; corrosive environments can shorten conductor life by a factor of 1.17 \cite{IEC2015}. Industrial conductor failure cases\cite{AusNet2025} further highlight the need to account for the risk of ETO and corrosion in grid operation.
Moreover, while DLR exhibits strong synergy with wind power, the correlation between WP and DLR forecasts can act as a source of risk amplification, not merely operational benefit.
However, this risk was not considered in \cite{Wallnerstrom2015} and \cite{Lee2022}. 
Prior work has attempted to address forecast uncertainty using copula-based chance-constraints \cite{Viafora2020} or distributionally robust formulations \cite{Wang2018}. 
Park et al. \cite{Park2018} extended this to stochastic unit commitment, showing that commitment decisions differ under SLR and DLR. 
Other works proposed risk-averse forecasts using quantile or super-quantile regression \cite{Kirilenko2021}.
In practice, FERC requires a quantile forecast value with a confidence level of 98\% \cite{FERC2024}. 
Recent work by \cite{Kim2025} develops such quantile forecasts using a graph convolutional LSTM model.
 
Despite these advances, existing models do not account for the actual costs of ETO nor provide a clear explanation on the effect of WP and DLR forecast error correlation.
Without quantifying ETO-induced depreciation, models remain mathematically robust but economically incomplete. 
Previous works have handled the effect of ETO, and some have expanded it to DLR, but no works have integrated ETO costs into decision making such as OPF or UC. 
Harvey's seminal work \cite{Harvey1972} modeled the effect of ETO on aluminum conductors through an empirical equation for tensile strength loss (LoTS) as a function of conductor temperature and elapsed time. 
This model has been widely adopted to model ETO, including Musilek et al. \cite{Musilek2012}, EPRI \cite{EPRI2005} and PJM \cite{PJM2022}. 
\cite{Musilek2012} analyzed the historical aging of a conductor using meteorological data. 
Moreover, \cite{Teh2019} modeled the risk of DLR using the LoTS model and conducted a sensitivity analysis. 
However, this research does not consider the actual operation and performs only a simple calculation based on one line. 
A fretting-based reliability analysis model \cite{Morozovska2020}, points out the significance of corrosion modeling with ETO, but relies on large assumptions and limited empirical support.
To the best of our knowledge, no existing optimization framework integrates realistic conductor health costs into grid operation.

Incorporating ETO into optimization requires embedding a conductor temperature estimation model. 
However, owing to the nonlinear and multimodal (affected by many weather components) current-temperature relationship, direct integration into optimization is impractical.
Previous studies have proposed conductor temperature models within OPF formulations \cite{Kirilenko2021,Ngoko2018}. 
While these works employ current–temperature proxies, they do not capture the consequences of ETO, leaving a major gap in the literature.

This paper addresses this gap by developing a day-ahead stochastic UC model that accounts for (i) DLR operation, (ii) correlated DLR and WP uncertainties, and (iii) the long-term costs of ETO.
Incorporating ETO costs is challenging due to the state dependency of conductor depreciation models and the lack of a closed-form solution to the loading-temperature relationship. To overcome these challenges, we introduce a piecewise linear conductor depreciation cost function and a linear line loading–temperature proxy.
The main contributions of this paper are as follows:
\begin{enumerate}
    \item \textit{Modeling of ETO effects}: We develop tractable models to capture the long-term effects of ETO, including (i) a robust and computationally efficient line loading–temperature proxy, and (ii) a risk-based piecewise conductor depreciation cost function.
    \item \textit{Conductor Health-Aware Unit Commitment (CHA-UC)}: Building on these models, we formulate a stochastic UC framework that internalizes conductor health costs in operational decision making, explicitly balancing conductor depreciation against system operating costs.
    \item \textit{Correlation risk characterization}: We characterize how the correlation between WP and DLR forecast errors affects ETO-induced depreciation, providing theoretical justification for incorporating WP correlation in DLR risk-aware operation.
    \item \textit{Validation on a realistic test system}: We evaluate the CHA-UC on the Texas 123-bus backbone test system (TX-123BT) under uncertainty. Results highlight differences in commitment decisions and demonstrate improved handling of DLR and WP risks.
\end{enumerate}

\section{Conductor cost depreciation model}\label{sec:CCDM}
This section quantifies the long-term effects of elevated conductor temperature operation (ETO) over the thermal limit and formulates a depreciation model suitable for optimization. 

\subsection{Loss of tensile strength (LoTS) for ACSR lines}\label{subsec:CHM}
Long-term ETO causes conductor degradation, measured as the loss of tensile strength (LoTS). 
For aluminum strands, LoTS (\%) is empirically modeled as shown in \cite{Harvey1972}:
\begin{align}\label{Eq:AlDeg}
    L^{\mathrm{Al}} = \gamma (100 - A\,t^{-\frac{25.4}{d}(0.001\tau - 0.095)}),
\end{align}
where $A=\min(100,\,134-0.24\tau)$, $\tau$ is conductor temperature (°C), $t$ is exposure time (h), and $d$ is conductor diameter (mm). $\gamma$, an environmental corrosivity factor, is introduced to reflect the effect of corrosive environments.
For aluminum conductor steel-reinforced (ACSR) lines, the corresponding LoTS is approximated as given in \cite{Musilek2012}:
\begin{align}\label{Eq:L_ACSR}
L^{\text{ACSR}} = 0.735\,L^{\text{Al}}.
\end{align}
The accumulated LoTS is obtained by integrating temperature exposure over time, enabling direct quantification of ETO-induced degradation. 

\subsection{Probability of Failure from LoTS}\label{subsec:PoF}
The consequence of ETO can be modeled through the increased probability of failure (PoF) of the Overhead conductor. Instantaneous \textit{a posteriori} PoF is used, reflecting the PoF given the survival until conductor LoTS $L^\text{ACSR}$:
\begin{align} \label{Eq:PoF}
\pi(L^\text{ACSR}) \;=\; \frac{f(L^\text{ACSR})}{\displaystyle \int_{L^\text{ACSR}}^{\infty} f(\ell)\, d\ell},
\end{align}
where $f(\cdot)$ is the probability density function (PDF) for modeling the failure. The use of \eqref{Eq:PoF} enables modeling failure probability considering the events leading up to $L^\text{ACSR}$ \cite{Teh2017}.

\subsection{Piecewise depreciation cost for overhead lines}\label{subsec:CC}
\begin{figure}
    \centering    
    \includegraphics[width=0.9\columnwidth]{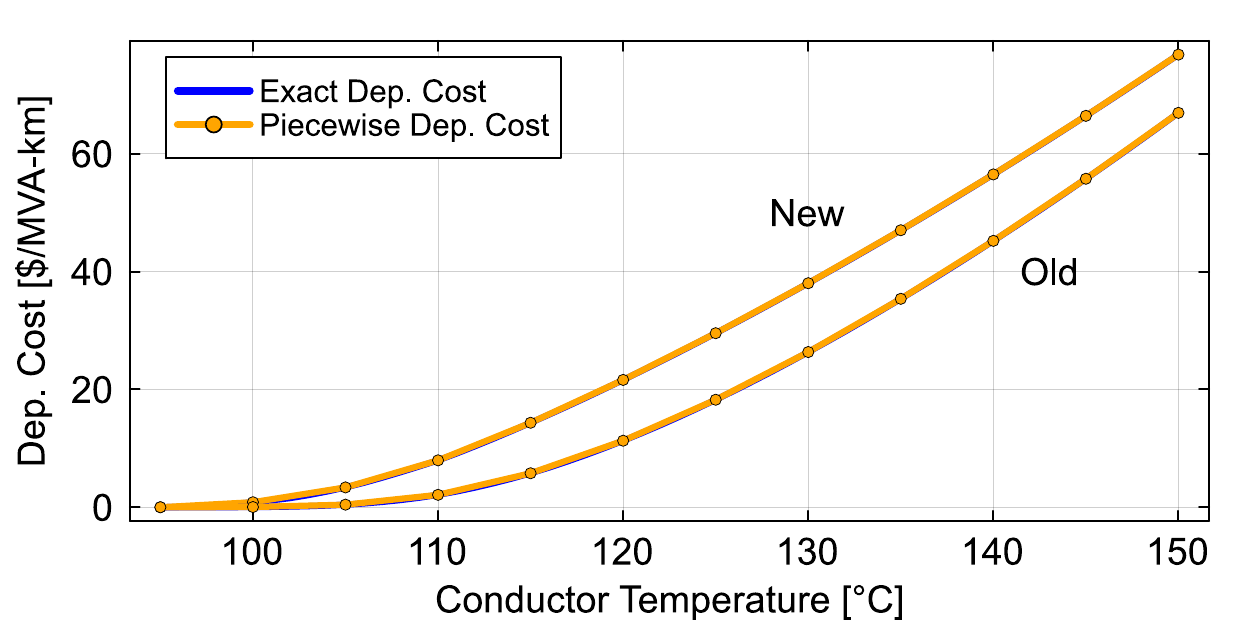}
    \caption{\small 
    Piecewise function of the incremental depreciation cost per one hour of ETO. Functions are drawn for each new and old ACSR Finch. The piecewise functions lay above the exact depreciation cost.}
    \label{fig:DepCost}
    \vspace{-2mm}
\end{figure}

The increased risk due to ETO is translated into economic depreciation. Utilities replace conductors when failure risk exceeds reconductoring cost \cite{AusNet2025}, yielding: 
\begin{align} \label{Eq:Risk}
\pi(L^\text{ACSR, End}_l)\cdot C^\text{Fail}_l - C^\text{Replace}_l = 0,
\end{align}
where $C^\text{Fail}_l$ is the cost of failure of line $l$ and $C^\text{Replace}_l =  B_l\cdot S_l\cdot D_l$. Here, $B_l$ is the cost factor of $l$ (\$/MVA-km), $S_l$ its capacity (MVA), and $D_l$ its length (km). 
Using the typical end-of-life threshold $L^\text{ACSR, End}_l=10$ \cite{Adomah2000, miso2021workshop}, this becomes
\begin{align} \label{Eq:DepCost1}
C^\text{Fail}_l \cdot \pi(L^\text{ACSR}_l) = \frac{C^\text{Replace}_l \cdot \pi(L^\text{ACSR}_l)}{\pi(10)}
\end{align}
By subtracting equation \eqref{Eq:DepCost1} for a certain timestep, the depreciation (or risk) cost due to incremental LoTS is then: 
\begin{align}\label{Eq:DepCost}
C^\text{Dep}_{l,t}(\tau) = C^\text{Replace}_l\cdot \frac{\pi(L^\text{ACSR}_{l,t}(\tau)) - \pi(L^\text{ACSR}_{l,t-1})}{\pi(10)},
\end{align}
where subscript $t$ denotes the time of interest. 
Because $C^\text{Dep}_t(\tau)$ depends on the conductor state, its direct use is limited to post-hoc evaluation. To embed depreciation costs into the daily UC problem, this study constructs conductor-specific convex piecewise approximations of the one-hour ETO cost as a function of temperature. 
Figure~\ref{fig:DepCost} compares the exact and piecewise depreciation costs for two lines; the old line exhibits lower incremental costs since conductor health becomes less sensitive to ETO exposure as LoTS accumulates.
\vspace{-2mm}
\subsection{Heat balance equation}\label{subsec:HBE}
The steady-state heat balance equation (HBE) \cite{IEEE738} is used to obtain conductor temperature $\tau$.
Since our focus is on hour-scale unit commitment, transients are neglected: conductor reaches 95\% of the steady-state temperature within approximately 42 minutes \cite{Park2018}. The steady-state HBE is:
\begin{align}
q_c + q_r = q_s + I^2 R(\tau_{\text{avg}}) \label{Eq:HBE} 
\end{align}
where $q_c$, $q_r$, and $q_s$ denote convective and radiative heat losses and solar heat gain, respectively. $I^2 R(\tau_{\text{avg}})$ is the Joule heating term.
Among these, wind speed (convection) has the strongest influence on the DLR, followed by ambient temperature \cite{Wallnerstrom2015}. 

\subsection{Linear temperature approximation}\label{subsec:LinTemp}
\begin{figure}
    \centering 
    \includegraphics[width=0.9\columnwidth]{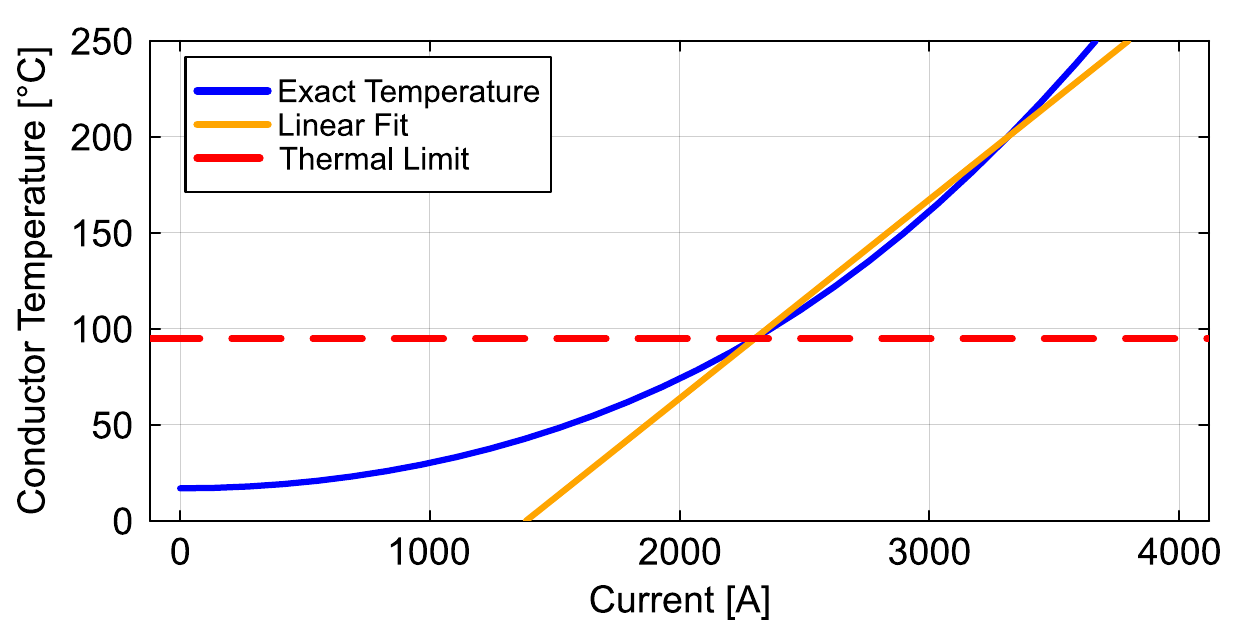}
    \caption{\small Illustration of a linear fit to the current-conductor temperature relationship, given a perpendicular wind speed of 2.7 m/s and ambient temperature of 9.4\textdegree C. Conductor temperature limit is set as 95 \textdegree C; in practice, it depends on the transmission owner \cite{Brown2024}.}
    \label{fig:LinearFit}
    \vspace{-4mm}
\end{figure}

The current-temperature relationship obtained from equation \eqref{Eq:HBE} cannot be incorporated straight to an optimization problem, since it lacks a closed-form solution. 
The current–temperature relationship is convex under realistic operating conditions,\footnote{Confirmed through parametric sweeps of ACSR Finch (current: 500–3500 A, wind speed: 0–8 m/s, ambient temperature: 0–45 °C, solar irradiance: 0-1200 W/$m^2$). Concave regions occur only under conditions beyond the safe operating range for ACSR ($\tau \ge 350$ °C, wind speed $\le 2$ m/s).}
making the linear proxy conservative and computationally efficient.
Figure \ref{fig:LinearFit} shows the linear fit of the HBE temperature curve.  
Within the emergency loading range (100–150\% of rated current \cite{ATC2012}), the linear proxy yields a MAE of 1.35\% and a maximum error of 1.92\%. 
Since the function is convex, the fit overestimates temperature, ensuring conservativeness. 
\allowdisplaybreaks

This loading-temperature relationship can be embedded in optimization problems including the DC power flow (PF) equation. The constraints are formulated as ($\forall l\in\mathcal{L},t \in\mathcal{T}$):
\begin{subequations}
\label{Eq:lintemp}
\begin{align}
&\tau_{l,t}\ge A_{l,t} \cdot p_{l,t,} + B_{l,t}, 
,\label{Eq:lintemp1}
\\
&\tau_{l,t}\ge -A_{l,t} \cdot p_{l,t} + B_{l,t},
\label{Eq:lintemp2}
\end{align}
\end{subequations}
where $p_{l,t}$ is the line power flow. 
Since voltage is assumed constant in the DC power flow, $A_{l,t}$ is scaled by $\frac{1}{\sqrt{3}V^\text{Nom}_l}$ from the current–temperature fit. 
The two inequalities account for bidirectional flow, ensuring $\tau_{l,t}$ is an upper bound of the actual conductor temperature. 

\allowdisplaybreaks
\section{Conductor health-aware unit commitment}\label{sec:CHAUC}

\begin{figure}[t]
    \centering    
    \includegraphics[width=\columnwidth]{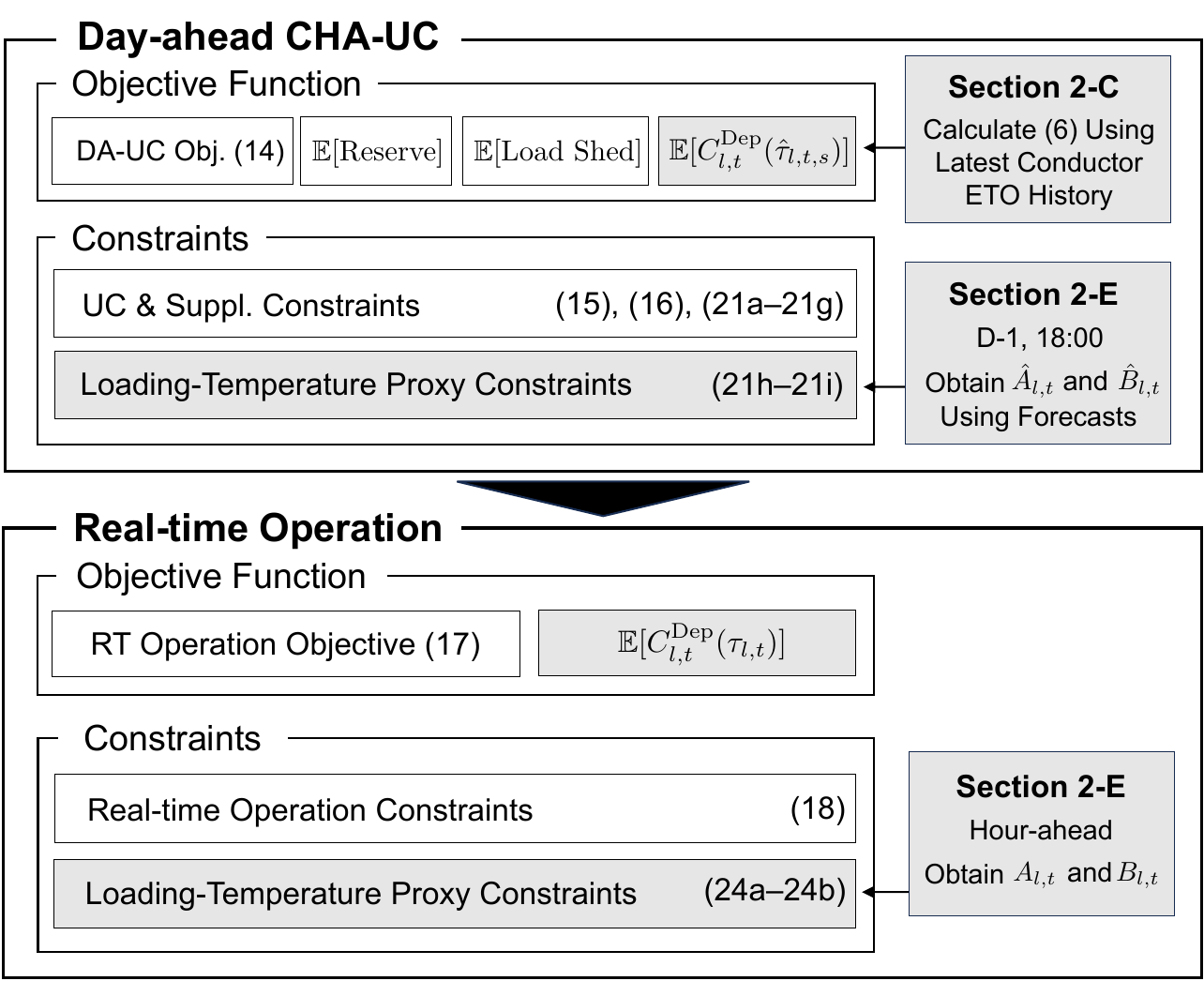}
    \caption{\small 
    Schematic of the CHA-UC optimization timeline. Processes that require actions outside optimization is colored with gray.}
    \label{fig:Schematic}
    \vspace{-4mm}
\end{figure}

Adopting DLR pushes line temperatures toward thermal limits, amplifying ETO risk in the presence of forecast errors. 
This section first characterizes the effect of correlation between WP and DLR forecast errors on ETO risk, motivating the explicit consideration of correlation in risk-averse DLR models.
Then we propose the CHA-UC framework, which minimizes the sum of expected conductor depreciation and system operating costs to balance reliability and economics.
The CHA-UC timeline is shown in Figure~\ref{fig:Schematic}. 

\subsection{Correlation of DLR and WP Forecast Errors}

This section characterizes how WP–DLR forecast error correlation affects ETO-induced conductor depreciation.

Since the depreciation cost is a convex function of thermal stress, we recall the following lemma from \cite{Shaked2007}.

\begin{lemma} \label{lem}
Let $X$, $Y$ be two random variables with the same mean. If $Y$ is more dispersed than 
$X$ in a mean-preserving spread, then for any non-decreasing convex function $g(\cdot)$,
\begin{equation}
\mathbb{E}[g(Y)] \ge \mathbb{E}[g(X)].
\end{equation}
\end{lemma}

Here, let the WP and DLR forecast errors be
\begin{align}
e_w &= \hat{P}^\text{Wind}_w - P^\text{Wind}_w, \,\,\,
e_l = \hat{\overline{F}}^\text{DLR}_l - \overline{F}^\text{DLR}_l,
\end{align}
and given $s_l=\mathrm{sign}(p_l),$ let the stochastic part of the thermal stress on line $l$ induced by wind farm $w$ be
\begin{equation}
\tilde{S}_{l,w} = A_l\bigl(e_l - s_l \mathrm{GSF}_{w,l}e_w\bigr),
\end{equation}
which captures the impact of forecast errors on line loading.

\begin{proposition}\label{prop1}
For a wind farm--line pair $(w,l)$, if $\mathrm{GSF}_{w,l} p_l < 0$, then increasing correlation $\rho_{w,l}$ between $e_w$ and $e_l$ increases the dispersion of $\tilde{S}_{l,w}$. If $\mathrm{GSF}_{w,l} p_l > 0$, then increasing correlation $\rho_{w,l}$ decreases the dispersion of $\tilde{S}_{l,w}$.
\end{proposition}

\begin{proof}
The proof is presented in Appendix \ref{appendixA}
\end{proof}

Combining \textbf{Proposition} \ref{prop1} with \textbf{Lemma} \ref{lem}, we obtain the following result on expected depreciation cost.

\begin{corollary}\label{cor2}
For fixed marginal distributions of $e_w$ and $e_l$, increasing positive correlation $\rho_{w,l}$ increases $\mathbb{E}[C_l^\mathrm{Dep}]$ if $\mathrm{GSF}_{w,l}p_l<0$, and decreases $\mathbb{E}[C_l^\mathrm{Dep}]$ if $\mathrm{GSF}_{w,l}p_l>0$.
\end{corollary}

\begin{proof}
The proof is presented in Appendix \ref{appendixB}
\end{proof}

Thus, the correlation between WP and DLR forecast errors exerts a systematic influence on line depreciation costs, depending on network topology and line flow conditions.
Wind is the dominant common driver of both WP and DLR, creating strong correlation between their forecast errors, especially for nearby assets.
Therefore, this effect should be accounted for in DLR operation, and the implementation of DLR on wind-connected lines. Notably, for typical WP tie lines, $\mathrm{GSF}_{w,l}p_l>0$ holds, so the correlation tends to hedge ETO risk because the wind farm is the only directly connected source.

\subsection{Baseline UC model}\label{subsec:BaseUC}
Before presenting the CHA-UC framework, the deterministic baseline model for DLR and SLR is introduced. Day-ahead UC and real-time operation is reflected in the model. The UC problem fixes commitment decisions day before the actual operation, using day-ahead forecasts and considering physical constraints of generators. The enforcement of UC constraints lead to a more comprehensive benchmark of results~\cite{Park2018}. 

The objective function of the day-ahead UC problem is:
\begin{align}
\label{Eq:baselineObj}
    \begin{split}
        &\min_{\Xi} \quad
            \sum_{t \in \mathcal{T}}\sum_{i \in \mathcal{I}}\bigg[ u_{i,t} \cdot C^{\text{P}}_i(g_{i,t}) + \sum_{o\in\mathcal{O}_i}\delta_{i,o,t} \cdot C^{\text{SU}}_{i,o}
        \\& \qquad \qquad \qquad \quad +u^\text{SD}_{i,t}\cdot C^{\text{SD}}_i \bigg] + \sum_{t \in \mathcal{T}}\sum_{b \in \mathcal{B}}C^{\text{VoLL}}p^{\text{shed}}_{b,t},
    \end{split}
\end{align}
which is the sum of production cost, startup cost and the shut down cost. $\Xi := \{\mathbf{g,p,u,}\,\boldsymbol{\theta,\delta}\}$ 
\allowdisplaybreaks

The day-ahead UC constraints, hereafter referred to as the UC constraints, are adapted from~\cite{Morales2013}:
\begin{subequations}
\label{Eq:baselineConstraints1}
\begin{align}
&\delta_{i,o,t} \le \sum^{T_{i,o+1}^{\text{SU}}-1}_{k = T^\text{SU}_{i,o}} u^\text{SD}_{i,t-i}, \quad \forall i \in \mathcal{I}, t\in [T_{i,o+1}^{\text{SU}},T], o \in [1,O_i), \label{Eq:segment}\\
&\sum_{o\in\mathcal{O}_i}\delta_{i,o,t} = u^\text{SU}_{i,t} 
\quad \forall i \in \mathcal{I} t \in \mathcal{T}, \label{Eq:segment2}\\
&\sum_{k= t-TU_i+1 }^t u^\text{SU}_{i,k} \le u_{i,t}, 
\quad\forall i \in \mathcal{I}, t\in [\text{TU}_i,T], \label{Eq:minSU}\\
&\sum_{k= t-TD_i+1 }^t u^\text{SD}_{i,k} \le 1- u_{i,t}, \quad
\forall i \in \mathcal{I}, t\in [\text{TD}_i,T],\label{Eq:minSD}\\
&u_{i,t}-u_{i,t-1} = u^\text{SU}_{i,t} - u^\text{SD}_{i,t}, \quad\forall i \in \mathcal{I}, t \in \mathcal{T},\label{Eq:logical}\\
&g_{i,t} \le (\overline{G}_i - \underline{G}_i)u_{i,t}- (\overline{G}_i-SU_i)u^\text{SU}_{i,t}, \,\,\, \forall i \in \mathcal{I}^1, t \in \mathcal{T}, \label{Eq:gen1}
\\
&g_{i,t} \le (\overline{G}_i - \underline{G}_i)u_{i,t}- (\overline{G}_i-SD_i)u^\text{SD}_{i,t+1}, \,\,\forall i \in \mathcal{I}^1, t \in \mathcal{T},\label{Eq:gen2}\\
\begin{split}
&g_{i,t} \le (\overline{G}_i - \underline{G}_i)u_{i,t}- (\overline{G}_i-SU_i)u^\text{SU}_{i,t}
\\ &  \qquad \qquad - (\overline{G}_i-SD_i)u^\text{SD}_{i,t+1}, \quad \forall i \in \mathcal{I}\setminus\mathcal{I}^1, t \in \mathcal{T},
\end{split} \label{Eq:gen3} \\
&g_{i,t} - g_{i,t-1} \le RU_i \quad\forall i \in \mathcal{I}, t \in \mathcal{T},\label{Eq:RU}
\\
& - g_{i,t} + g_{i,t-1} \le RD_i \quad\forall i \in \mathcal{I}, t \in \mathcal{T}.\label{Eq:RD} 
\end{align}
\end{subequations}
Startup and shutdown costs are modeled by \eqref{Eq:segment}–\eqref{Eq:segment2}, and minimum up/down times by \eqref{Eq:minSU}–\eqref{Eq:minSD}. 
The logical constraint \eqref{Eq:logical} links the binary variables, while \eqref{Eq:gen1}–\eqref{Eq:gen3} enforce generation limits 
($\mathcal{I}^1$ denotes units with $\text{TU}_i=1$). 
Ramping limits are imposed by \eqref{Eq:RU} and \eqref{Eq:RD}.

Supplementary constraints for the day-ahead UC are:
\begin{subequations}
\label{Eq:baselineConstraints2}
\begin{align}
&p_{l,t} = \tfrac{1}{X_l}\bigl(\theta_{s(l),t}-\theta_{r(l),t}\bigr), 
\quad\forall l\in\mathcal{L}, t \in \mathcal{T},\label{Eq:DCPF}\\
\begin{split}
&\sum_{l\mid s(l)=b}\hspace{-1.5mm}p_{l,t} 
-\hspace{-3mm}\sum_{l\mid r(l)=b}\hspace{-1.5mm}p_{l,t}+\sum_{i\in\mathcal{I}_b}g_{i,t}
 + \hspace{-3mm}\sum_{w\in \mathcal{W}_b} (\hat{P}^{\rm wind}_{w,t}- p^{\text{cur}}_{w,t})
\\
&=
P^\text{d}_{b,t} - p^\text{shed}_{b,t},
\quad\forall b\in\mathcal{B},t\in \mathcal{T},
\end{split}\label{Eq:Pbalance}\\
&p^{\text{cur}}_{w,t}\le\hat{P}^{\rm wind}_{w,t}, \quad \forall w\in\mathcal{W}, t\in\mathcal{T} \label{Eq:Pcur}\\
&r^{\text{avail},+}_{i,t} \le R^{\mathrm{RU}}_{i}\,u_{i,t},\quad
\forall i\in\mathcal I, t\in\mathcal T, \label{eq:reg-ru} \\
&r^{\text{avail},+}_{i,t} \le \overline P_i - g_{i,t},
\quad\forall i\in\mathcal I, t\in\mathcal T, \label{eq:reg-headroom}
\\
&\sum_{i\in\mathcal I} r^{\text{avail},+}_{i,t} \;\ge\; R^{\text{min},+}_t,
\quad \forall  t\in\mathcal T, \label{eq:reg-requirement}\\
&\underline{F}^{\rm SLR}_l \,\le\, p_{l,t} \,\le\, \overline{F}^{\rm SLR}_l,
\quad\forall l\in\mathcal{L}^\text{SLR},t\in\mathcal{T},\label{Eq:SLR}\\
&K_{l,t} \hat{\underline{F}}^{\rm DLR}_{l,t} \,\le\, p_{l,t} \,\le\, K_{l,t}\hat{\overline{F}}^{\rm DLR}_{l,t},
\quad\forall l\in\mathcal{L}^\text{DLR},t\in\mathcal{T}.\label{Eq:DLR}
\end{align}
\end{subequations}
The network is modeled with DC power flow~\eqref{Eq:DCPF}, and nodal balance is enforced by~\eqref{Eq:Pbalance}. 
Equations~\eqref{Eq:Pcur}–\eqref{eq:reg-requirement} impose wind curtailment and reserve requirements. 
Angle and flow envelopes~\eqref{Eq:SLR}–\eqref{Eq:DLR} ensure grid security, where $K_{l,t}\!\in\![0,1]$ is the operator-defined DLR margin.

Once the day-ahead dispatch is fixed, the real-time operation minimizes the additional cost of balancing actual wind generation and weather realizations:
\begin{align}
\begin{split}
 &\min_{\Xi} \,\,
\sum_{t \in \mathcal{T}}\sum_{i \in \mathcal{I}}\bigg[C^{\text{P,2}}_i(g^{\text{adj}}_{i,t})^2 + C_i^{\text{r}+}r^{+}_{i,t} + C_i^{\text{r}-}r^{-}_{i,t} \bigg] \\
 &\qquad + \sum_{t \in \mathcal{T}}\sum_{b \in \mathcal{B}}C^{\text{VoLL}}p^{\text{shed,adj}}_{b,t},
\end{split}\label{Eq:BaselineRTObj}
\end{align}
where $\Xi := \{\mathbf{g}^\mathrm{adj}, \mathbf{p}^\mathrm{adj},\mathbf{r}^\mathrm{+,-},\boldsymbol{\theta}^\mathrm{adj} \}$. 
Reserve activations are penalized through $C_i^{\mathrm{r}\pm}$, 
and the quadratic term captures nonlinear production costs. 
This objective represents the operator’s total adjustment cost in real time.
The constraints of the operation stage are as follows:
\begin{subequations}
\label{Eq:baselineConstraints3}
\begin{align}
& g^\text{adj}_{i,t} = \overline{g}_{i,t} + r^+_{i,t}-r^-_{i,t},
\quad\forall i \in \mathcal{I}, t\in\mathcal{T}, \label{Eq:baseRTgen} \\
& 0\le r^+_{i,t} \le \overline{u}_{i,t} \cdot G^{\text{RU}}_{i},
\quad\forall i \in \mathcal{I}, t\in\mathcal{T},\label{Eq:baseRTup}\\
& 0\le r^-_{i,t} \le \overline{}{u}_{i,t} \cdot G^{\text{RD}}_{i},
\quad\forall i \in \mathcal{I}, t\in\mathcal{T},\label{Eq:baseRTdn}\\
&  \underline{G}_i\le g^\text{adj}_{i,t} \le \overline{G}_{i},
\quad\forall i \in \mathcal{I}, t\in\mathcal{T},\label{Eq:baseRTlim}\\
&p^\text{adj}_{l,t} = \tfrac{1}{X_l}\bigl(\theta^\text{adj}_{s(l),t}-\theta^\text{adj}_{r(l),t}\bigr), 
\quad\forall l\in\mathcal{L}, t \in \mathcal{T},\label{Eq:baseRTpf}\\
\begin{split}
&\sum_{l\mid s(l)=b}p^\text{adj}_{l,t} 
-\sum_{l\mid r(l)=b}p^\text{adj}_{l,t}
+\sum_{i\in\mathcal{I}_b}g^\text{adj}_{i,t}+ \sum_{w\in \mathcal{W}_b} (P^{\rm wind}_{w,t}
\\& \qquad - p^{\text{cur,adj}}_{w,t})
=
P^\text{d}_{b,t} - p^\text{shed,adj}_{b,t},
\quad\forall b\in\mathcal{B},t\in \mathcal{T},
\end{split}\label{Eq:baseRTpbal}
\\
&p^{\text{cur,adj}}_{w,t}\le  P^{\rm wind}_{w,t}, \quad \forall w\in\mathcal{W}, t\in\mathcal{T}.\label{Eq:baseRTcur}
\end{align}
\end{subequations}
The real-time operation balances supply and demand under fixed commitment decisions by activating reserves and adjusting wind curtailment and load shedding. 
Day-ahead dispatch is corrected through up- and down-reserve activation as defined in \eqref{Eq:baseRTgen}, subject to ramping, generation, and commitment constraints \eqref{Eq:baseRTup}–\eqref{Eq:baseRTlim}. 
The real-time power balance \eqref{Eq:baseRTpbal} is updated to reflect actual wind generation and the adjusted power flows, generation, and curtailment.

\subsection{Conductor health-aware unit commitment}\label{subsec:CHAUC2}

Given the linear fit and the piecewise depreciation cost function in Section~\ref{sec:CCDM}, the objective function of the CHA-UC day-ahead problem is formulated:
\begin{align}
&\min_{\Xi} \quad
\sum_{t \in \mathcal{T}}\sum_{i \in \mathcal{I}}\bigg[
C^{\text{P}}_i(g_{i,t}) u_{i,t}
+ \sum_{o\in\mathcal{O}_i} C^{\text{SU}}_{i,o} \delta_{i,o,t}
+ C^{\text{SD}}_i u^\text{SD}_{i,t}
\bigg]
\notag\\
&
+ \sum_{t \in \mathcal{T}}\sum_{b \in \mathcal{B}}
C^{\text{VoLL}}p^{\text{shed}}_{b,t}
+\sum_{s \in \mathcal{S}} \mathrm{Pr}_s \sum_{t\in\mathcal{T}} \bigg[
\underbrace{\sum_{l\in\mathcal{L}^{\text{DLR}}} C^{\text{Dep}}_{l,t} (\hat{\tau}_{l,t,s})}_{\mathbb{E}(\text{Depreciation cost})}
\notag\\
&
+ \underbrace{\sum_{b \in \mathcal{B}} C^{\text{VoLL}}\hat{p}^\text{shed,adj}_{b,t,s}}_{\mathbb{E}(\text{Load shedding cost})}
+ \underbrace{\sum_{i \in \mathcal{I}}
\left[C^{\text{r+}}\hat{r}^\text{+}_{i,t,s} + C^{\text{r-}}\hat{r}^\text{-}_{i,t,s}\right]}_{\mathbb{E}(\text{Reserve activation cost})}
\bigg].
\label{Eq:CHAObj}
\end{align}

The objective includes the operation cost from~\eqref{Eq:baselineObj} and the stochastic costs: the expected conductor depreciation cost, load shed costs and reserve activation costs. Here, $C^{\text{Dep}}_{l,t}$ is implemented using \textbf{SOS2} constraints. All terms including the depreciation cost in the objective functions uses monetary units (\$), which eliminates the need for any additional scaling or conversion coefficient, simplifying the optimization and improving interpretability.
The decision variables are identical to the baseline. 
The sample average approximation was used to calculate the expected value. 

CHA-UC is constrained by the UC constraints and the DC power flow constraints from the baseline.
\begin{subequations}
\label{Eq:CHAConstraints_1}
\begin{align}
&\eqref{Eq:segment} - \eqref{Eq:RD}  &&\text{UC Constraints},\\
&\eqref{Eq:DCPF} - \eqref{Eq:DLR}  &&\text{Suppl. Constraints}.
\end{align}
\end{subequations}

Additional scenario-based stochastic constraints are as follows ($\forall s \in \mathcal{S}$): 
\begin{subequations}
\label{Eq:CHAConstraints_2}
\begin{align}
&\hat{g}^{\text{adj}}_{i,t,s} = g_{i,t} + \hat{r}^+_{i,t,s} - \hat{r}^-_{i,t,s}, 
\quad \forall i \in \mathcal{I}, t \in\mathcal{T}, \label{Eq:CHAgen}\\
& 0\le \hat{r}^+_{i,t,s} \le G^{\text{RU}}_{i},
\quad\forall i \in \mathcal{I}, t\in\mathcal{T},\label{Eq:CHAup}\\
& 0\le \hat{r}^-_{i,t,s} \le G^{\text{RD}}_{i},
\quad\forall i \in \mathcal{I}, t\in\mathcal{T}, \label{Eq:CHAdn}\\
& \underline{G}_i\le \hat{g}^\text{adj}_{i,t,s} \le \overline{G}_{i},
\quad\forall i \in \mathcal{I}, t\in\mathcal{T}, \label{Eq:CHAgenlim}\\
&\hat{p}^\text{adj}_{l,t,s} = \tfrac{1}{X_l}\bigl(\hat{\theta}_{s(l),t,s}^\text{adj}-\hat{\theta}_{r(l),t,s}^\text{adj}\bigr), 
\quad\forall l\in\mathcal{L}, t \in \mathcal{T}, \label{Eq:CHAdc}\\
\begin{split}
&\sum_{l\mid s(l)=b}\hspace{-1.5mm}\hat{p}^\text{adj}_{l,t,s} 
-\sum_{l\mid r(l)=b}\hspace{-1.5mm}\hat{p}^\text{adj}_{l,t,s}
+\sum_{i\in\mathcal{I}_b}\hat{g}^\text{adj}_{i,t,s}
+ \hspace{-1.5mm}\sum_{w\in \mathcal{W}_b} \hspace{-1.5mm}(\hat{P}^{\rm wind}_{w,t} 
\\ &  - \hat{p}^{\text{cur,adj}}_{w,t,s} - \omega_{w,t,s} )
=
P^\text{d}_{b,t} -\hat{p}^\text{shed,adj}_{b,t,s},
\quad\forall b\in\mathcal{B},t\in \mathcal{T}, 
\end{split} \hspace{-3mm}\label{Eq:CHApbal}\\
&\hat{p}^{\text{cur,adj}}_{w,t,s}\le \hat{P}^{\rm wind}_{w,t} - \omega_{w,t,s}, \quad\forall w\in\mathcal{W}, t\in\mathcal{T},\label{Eq:CHApcur}\\
&\hat{\tau}_{l,t,s}\ge \hat{A}_{l,t} (\hat{p}^{\text{adj}}_{l,t,s}+\xi_{l,t,s}) + \hat{B}_{l,t}, 
~\forall l\in\mathcal{L}^{\text{DLR}},t \in\mathcal{T},\label{Eq:CHAtemp1}\\
&\hat{\tau}_{l,t,s}\ge -\hat{A}_{l,t} (\hat{p}^{\text{adj}}_{l,t,s}-\xi_{l,t,s}) + \hat{B}_{l,t}, 
~\forall l\in\mathcal{L}^{\text{DLR}},t \in\mathcal{T}.\hspace{-1mm}\label{Eq:CHAtemp2}
\end{align}
\end{subequations}

The surrogate of the ``wait-and-see'' operation stage is formed to reflect the stochasticity in the ``here-and-now'' day-ahead UC. The surrogate corresponding to \eqref{Eq:baseRTgen}\textendash\eqref{Eq:baseRTcur} are \eqref{Eq:CHAgen}\textendash\eqref{Eq:CHApcur}, respectively. They all represent the same operational constraints, but the latter are not actual; they represent the constraints with scenario realizations. 
Random variables regarding each scenario, $\omega_{w,t,s}$ and $\xi_{l,t,s}$ represent the wind forecast error and DLR forecast error.
Equation \eqref{Eq:CHAtemp1} and \eqref{Eq:CHAtemp2} imposes a lower bound of $\tau_{l,t,s}$ using the linear conductor-temperature approximation \eqref{Eq:lintemp}. $\tau_{l,t,s}$ will be fixed to the lower bound as $C^\text{Dep}_{l,t}$ is a non-decreasing convex function, working as a proxy of conductor temperature. This linear approximation offers two advantages: it represents DLR forecast errors as equivalent flow deviations and handles bidirectional line flow efficiently.

The operation stage of a CHA-UC framework include an additional conductor depreciation cost term in the model. Other parts of the optimization problem are identical. The objective function of the CHA-UC real-time operation process is as follows. 
\begin{align}
\begin{split}
 &\min_{\Xi} \quad
\sum_{t \in \mathcal{T}}\sum_{i \in \mathcal{I}}\bigg[C^{\text{P,2}}_i(g^{\text{adj}}_{i,t})^2 + C_i^{\text{r}+}r^{+}_{i,t} + C_i^{\text{r}-}r^{-}_{i,t} \bigg] \\
 &\quad\,\,\,\,\,\, + \sum_{t \in \mathcal{T}}\sum_{b \in \mathcal{B}}C^{\text{VoLL}}p^{\text{shed,adj}}_{b,t}+ \underbrace{\sum_{t\in\mathcal{T}}\sum_{l \in \mathcal{L}^\text{DLR}}C^{\text{Dep}}_{l, t} (\tau_{l,t})}_{\text{Depreciation cost}}.
 \end{split}
\label{Eq:CHARTObj}
\end{align}

The constraints identical to the baseline model are:
\label{Eq:CHARTConstraints_1}
\begin{align}
&\eqref{Eq:baseRTgen} - \eqref{Eq:baseRTcur}  &&\text{Real-time operation},
\end{align}

Additional conductor temperature related constraints are:
\begin{subequations}
\label{Eq:CHARTConstraints_2}
\begin{align}
&\tau_{l,t}\ge A_{l,t} \cdot p^{\text{adj}}_{l,t,} + B_{l,t}, 
\quad\forall l\in\mathcal{L}^{\text{DLR}},t \in\mathcal{T},\label{Eq:CHARTtemp1}
\\
&\tau_{l,t}\ge -A_{l,t} \cdot p^{\text{adj}}_{l,t} + B_{l,t}, 
\quad\forall l\in\mathcal{L}^{\text{DLR}},t \in\mathcal{T}.\label{Eq:CHARTtemp2}
\end{align}
\end{subequations}

 Analogous to equations (\ref{Eq:CHAtemp1})-(\ref{Eq:CHAtemp2}), equations (\ref{Eq:CHARTtemp1})-(\ref{Eq:CHARTtemp2}) enforce $\tau_{l,t}$ to reflect the conductor temperature before the operation stage.

\begin{table*}[t]
  \centering
  \scriptsize  
  \vspace{-2.0mm}
  \captionsetup{justification=centering, labelsep=period, font=footnotesize, textfont=sc}
  \setlength{\tabcolsep}{3.5pt}
  \caption{\small Summarized results of Case Study I and Case Study II updated from the CSV files.
  The best result for each cost is in \textbf{bold}.}
  \label{tab:exp_compare_grouped2}
  \begin{tabular}{cc | ccccc | c | ccccc | c}
    \toprule
    & & \multicolumn{6}{c|}{\textbf{Case Study\,1} (no WP error)} & \multicolumn{6}{c}{\textbf{Case Study\,2} (with WP error)} \\
    & & \multicolumn{1}{c|}{Day-ahead} & \multicolumn{2}{c|}{Real-time} & \multicolumn{1}{c|}{Post-hoc} & \multicolumn{2}{c|}{DA+RT}
    & \multicolumn{1}{c|}{Day-ahead} & \multicolumn{2}{c|}{Real-time} & \multicolumn{1}{c|}{Post-hoc} & \multicolumn{2}{c}{DA+RT} \\
    \multicolumn{1}{c}{\textbf{Season}} &
    \multicolumn{1}{c|}{\textbf{Method}} &
    \multicolumn{1}{c|}{\textbf{Day Ahead}} &
    \textbf{Reserve} &
    \multicolumn{1}{c|}{\textbf{Load Shed}} &
    \multicolumn{1}{c|}{\textbf{Dep. Cost}} &
    \textbf{Curt.} &
    \textbf{Total} &
    \multicolumn{1}{c|}{\textbf{Day Ahead}} &
    \textbf{Reserve} &
    \multicolumn{1}{c|}{\textbf{Load Shed}} &
    \multicolumn{1}{c|}{\textbf{Dep. Cost}} &
    \textbf{Curt.} &
    \multicolumn{1}{c}{\textbf{Total}} \\

    & & \multicolumn{1}{c|}{[M\$/day]} &
    [M\$/day] & \multicolumn{1}{c|}{[M\$/day]} & \multicolumn{1}{c|}{[M\$/day]} & [MW/day] &
    \multicolumn{1}{c|}{[M\$/day]} &
    \multicolumn{1}{c|}{[M\$/day]} &
    [M\$/day] & \multicolumn{1}{c|}{[M\$/day]} & \multicolumn{1}{c|}{[M\$/day]} & [MW/day] &
    \multicolumn{1}{c}{[M\$/day]} \\
    \midrule \midrule

Spring & CHA & 14.29 & 0.08 & -- & 0.22 & 11.30 & \textbf{14.60}
              & 14.68 & \textbf{1.89} & \textbf{1.35} & \textbf{0.66} & 96.8 & \textbf{18.58} \\
       & DLR & \textbf{14.16} & -- & -- & 1.23 & \textbf{7.6} & 15.39
              & \textbf{13.73} & 2.71 & 3.89 & 11.72 & \textbf{66.2} & 32.05 \\
       & SLR & 14.57 & -- & -- & \textbf{0.13} & 92.7 & 14.70
              & 14.15 & 2.28 & 2.83 & 8.80 & 85.9 & 28.06 \\
       & QRF & 14.25 & -- & -- & 0.72 & 29.2 & 14.97
              & 13.81 & 2.63 & 3.75 & 8.67 & 74.1 & 28.86 \\
\midrule

Summer & CHA & 29.96 & 0.12 & -- & \textbf{0.12} & 14.3 & \textbf{30.21}
              & 29.77 & \textbf{2.92} & \textbf{2.37} & \textbf{0.28} & 54.3 & \textbf{35.34} \\
       & DLR & \textbf{29.63} & -- & -- & 0.97 & \textbf{9.8} & 30.60
              & \textbf{28.48} & 4.11 & 5.44 & 10.96 & 28.9 & 49.00 \\
       & SLR & 29.96 & -- & -- & 0.46 & 56.5 & 30.42
              & 28.84 & 3.67 & 4.29 & 8.46 & 34.6 & 45.26 \\
       & QRF & 30.35 & -- & -- & 0.19 & 24.1 & 30.55
              & 29.17 & 3.93 & 4.89 & 7.44 & \textbf{28.1} & 45.43 \\
\midrule

Fall   & CHA & 22.08 & 0.06 & -- & \textbf{0.01} & 26.7 & \textbf{22.15}
              & 21.99 & \textbf{2.56} & \textbf{1.30} & \textbf{0.16} & 83.8 & \textbf{26.01} \\
       & DLR & \textbf{21.85} & -- & -- & 0.48 & \textbf{24.8} & 22.33
              & \textbf{20.89} & 3.66 & 3.48 & 12.62 & 46.9 & 40.65 \\
       & SLR & 22.26 & -- & -- & 0.03 & 126.7 & 22.30
              & 21.28 & 3.26 & 2.80 & 7.32 & 72.9 & 34.66 \\
       & QRF & 22.01 & -- & -- & 0.07 & 47.6 & \textbf{22.08}
              & 21.04 & 3.57 & 3.38 & 8.60 & \textbf{38.9} & 36.58 \\
\midrule

Winter & CHA & 18.50 & 0.05 & -- & 0.13 & 6.7 & \textbf{18.68}
              & 18.63 & \textbf{2.06} & \textbf{0.70} & \textbf{0.14} & 51.5 & \textbf{21.53} \\
       & DLR & \textbf{18.39} & -- & -- & 0.67 & \textbf{4.4} & 19.06
              & \textbf{17.72} & 3.00 & 3.01 & 8.94 & \textbf{41.7} & 32.67 \\
       & SLR & 18.86 & -- & -- & \textbf{0.00} & 53.5 & 18.86
              & 18.15 & 2.62 & 2.02 & 5.30 & 59.8 & 28.08 \\
       & QRF & 18.44 & -- & -- & 0.30 & \textbf{4.4} & 18.74
              & 17.75 & 2.98 & 2.76 & 7.46 & 44.5 & 30.95 \\
\midrule

Annual & CHA & 21.23 & 0.08 & -- & \textbf{0.12} & 14.8 & \textbf{21.44}
              & 21.29 & \textbf{2.36} & \textbf{1.44} & \textbf{0.31} & 71.8 & \textbf{25.41} \\
       & DLR & \textbf{21.04} & -- & -- & 0.84 & \textbf{11.7} & 21.87
              & \textbf{20.23} & 3.38 & 3.97 & 11.08 & \textbf{46.0} & 38.65 \\
       & SLR & 21.44 & -- & -- & 0.16 & 82.6 & 21.60
              & 20.63 & 2.96 & 3.00 & 7.49 & 63.3 & 34.08 \\
       & QRF & 21.29 & -- & -- & 0.32 & 26.5 & 21.61
              & 20.47 & 3.28 & 3.71 & 8.04 & 46.4 & 35.50 \\
\bottomrule
\end{tabular}
\vspace{-2mm}
\end{table*}

\section{Case Study}\label{sec:CS}

Two case studies are performed on the synthetic Texas 123-bus backbone transmission system (TX-123BT)~\cite{Lu2025} using 2021-2022 weather data. 
Case Study I runs an annual simulation with uncertain DLR forecasts (with errors) and perfect wind power (WP) forecasts (no errors) to illustrate CHA-UC. 
Case Study II evaluates CHA-UC in a more realistic setting with uncertainty in both WP and DLR forecasts to analyze their correlation effects.
\allowdisplaybreaks
\begin{figure*}[t]
  \centering%
    \subfloat[DLR, SLR and CHA-UC ]{%
    \includegraphics[width=\textwidth]{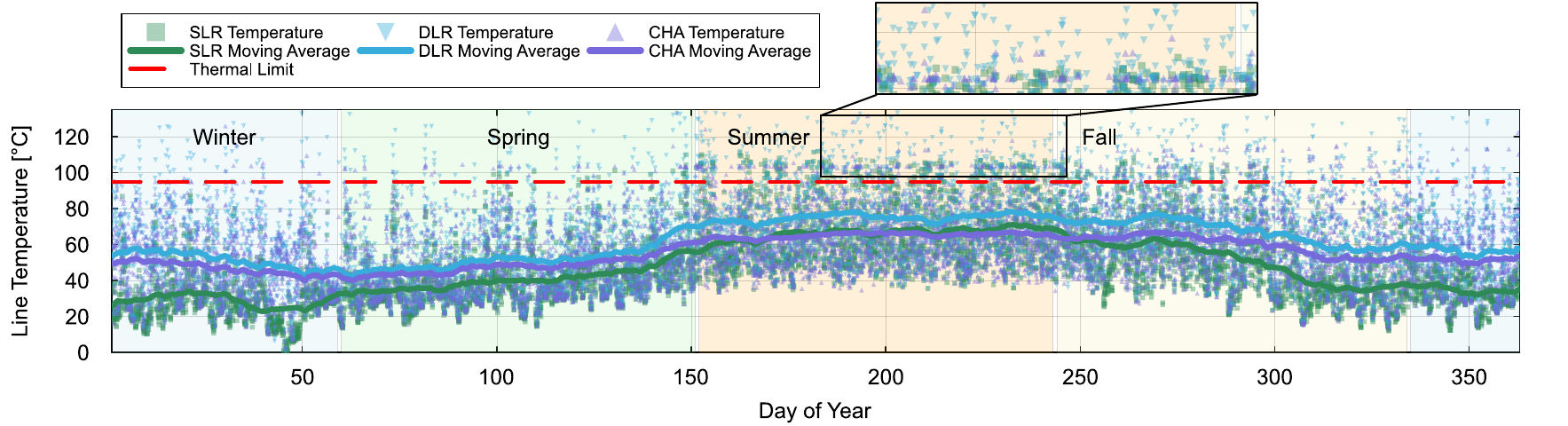}%
    \label{fig:temp_year}%
  }\\
    \subfloat[QRF and CHA-UC ]{%
    \includegraphics[width=\textwidth]{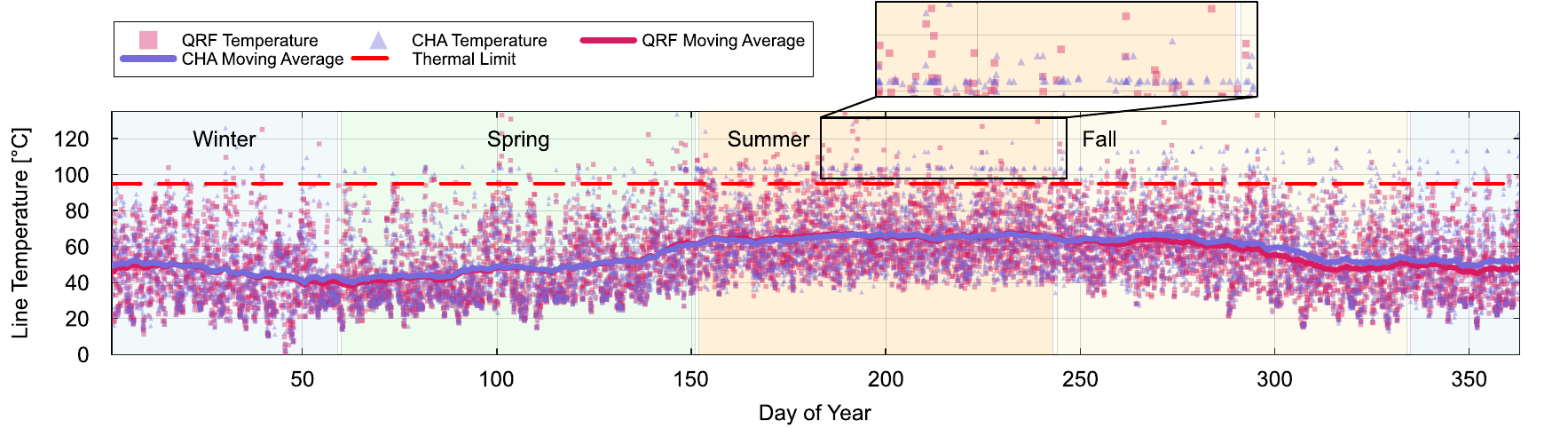}%
    \label{fig:temp_year_qrf}%
  }
  \\[0.2ex]
  \caption{\small (Case Study I) Annual scatter plot of temperatures of line 192 (Line with biggest temperature difference) obtained through post-hoc evaluation. 20-day moving average (MA) lines are drawn above. (a) The temperatures of DLR, SLR and CHA-UC are compared. A zoom-in of thermal overload temperatures is provided. (b) The temperatures of QRF and CHA-UC are compared. }
  \label{fig:temp_dist_seasons_year}
  \vspace{-2mm}
\end{figure*}
\subsection{Preliminaries} \label{sec:prelim}
\subsubsection{Simulation Settings} \label{sec:Setting}

In all case studies, a set of 21 lines was selected for DLR application\footnote{Lines were screened by solving an annual DCOPF with SLR and selecting those with the highest dual values of the line-flow constraint.}.
Each 345~kV line was modeled as ACSR Finch, and ampacity was adjusted to fit the TX-123BT system through a line multiplier. 
All optimization models were implemented in Julia/JuMP~\cite{Lubin2023} and solved with Gurobi. 
The LoTS limit was set to 10\% (Section~\ref{subsec:CC}). Corrosivity factor $\gamma$ was obtained from mapping steel corrosivity zone C2 and under to 1, C3 and above to 1.17, obtained from \cite{IEC2015, EngineeringDirector}.
The line cost factor $B_l =$\$936/MVA-km was obtained by excluding tower and loss components from a 400 kV, 3190 MVA case study \cite{Parsons2012}.
The PDF modeling line failure $f(\cdot)$ was modeled as a normal distribution following $\mu = 0.1, \sigma =0.05$ \cite{Teh2019}.
The conductor thermal limit was set to 95\textdegree C. The loading-temperature linear fit was sampled from 95 \textdegree C and 150 \textdegree C, and the piecewise depreciation cost function was sampled every 10 \textdegree C from 95 \textdegree C.
Conductor health was initialized to fit a mean LoTS rate of 1\% (a sensible rate from the results of \cite{Adomah2000}) by creating synthetic historical ETO exposures from a normal distribution. The historical ETO exposure temperatures are passed on between different horizons to update the conductor state properly.
For DLR security margin, line limits were constrained as 
$K_{l,t}\hat{\overline{F}}^{\mathrm{DLR}}_{l,t}=\max(0.8\hat{\overline{F}}^{\mathrm{DLR}}_{l,t},\,\underline{F}^{\mathrm{SLR}}_l)$.
The load profile of 2021 given in the TX-123BT test system was used.
The value of lost load, $C^\text{VoLL}$, is assumed to be \$3500/MWh following MISO before 2025/05 \cite{Kirschen2018}. 
Reserve activation costs followed \cite{Dvorkin2025}: $C_i^{\mathrm{r}+}$ equals three times, and $C_i^{\mathrm{r}-}$ equals 0.5~times, the linear cost coefficient of generator~$i$.

\subsubsection{Data Collection} \label{sec:Pre}
Weather data for 2021–2022 were obtained from NOAA’s HRRR model \cite{Benjamin2016}, with 3 km spatial and 1 h temporal resolution. Forecasts issued at 18{:}00 on the previous day were used for day-ahead inputs, and analysis data were treated as actual conditions. The 2022 data were used to generate stochastic scenarios, while the 2021 data were used for annual simulations. WP profiles were derived from wind speeds using a standard power curve \cite{Giorsetto1983} with hub-height correction, assuming uniform turbine specifications. For each line, the hourly minimum DLR among intersecting 3$\times$3 km grid cells was selected. HBE parameters followed IEEE \cite{IEEE738}, with solar heat gain fixed at a worst-case Texas condition.
\allowdisplaybreaks

\subsubsection{Scenario Generation for CHA-UC}\label{subsec:Scen}

To generate representative WP and DLR forecast-error scenarios while preserving their correlation, a daily trajectory dataset of 24-hour forecast errors is built. Historically extreme trajectories are first seeded, then a greedy max–min selection adds diverse trajectories. Scenario probabilities are then assigned using nearest-neighbor weighting with penalties on large deviations, yielding a scenario set that reflects the empirical error distribution.
A total of 20 scenarios were generated.
The statistic and correlation structure were well preserved: the mean, standard deviation and correlation varied by 0.7\%, 8\% and 28\% (Empirical 0.037 vs. Scenario 0.052), respectively.
\begin{figure*}[htbp]
  \centering
  \subfloat[Average of $u^\text{DLR Baseline}$ ]{%
    \includegraphics[width=0.9\columnwidth]{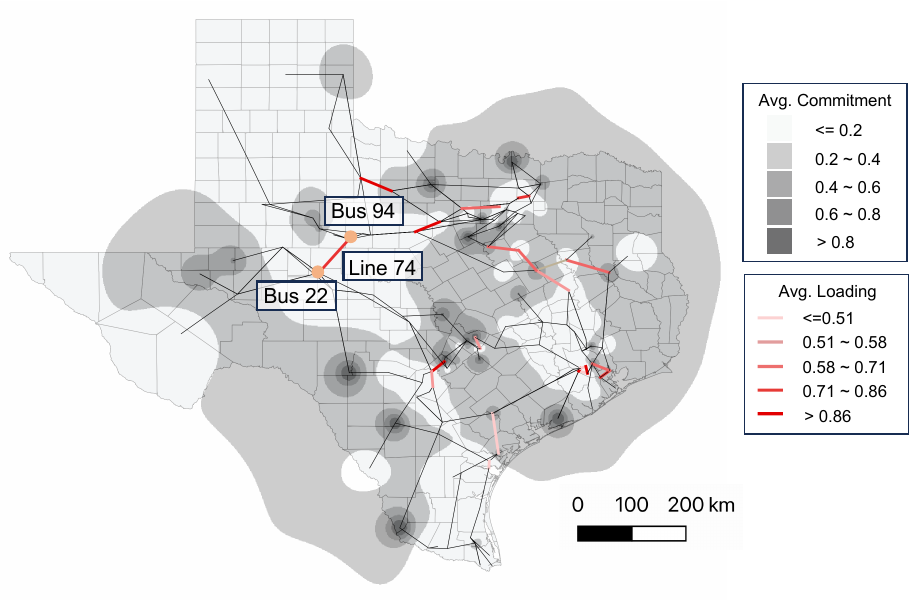}%
    \label{fig:QGIS_u}%
  }
  \subfloat[Average of $u^\text{CHA-UC}-u^\text{DLR Baseline}$]{%
    \includegraphics[width=0.9\columnwidth]{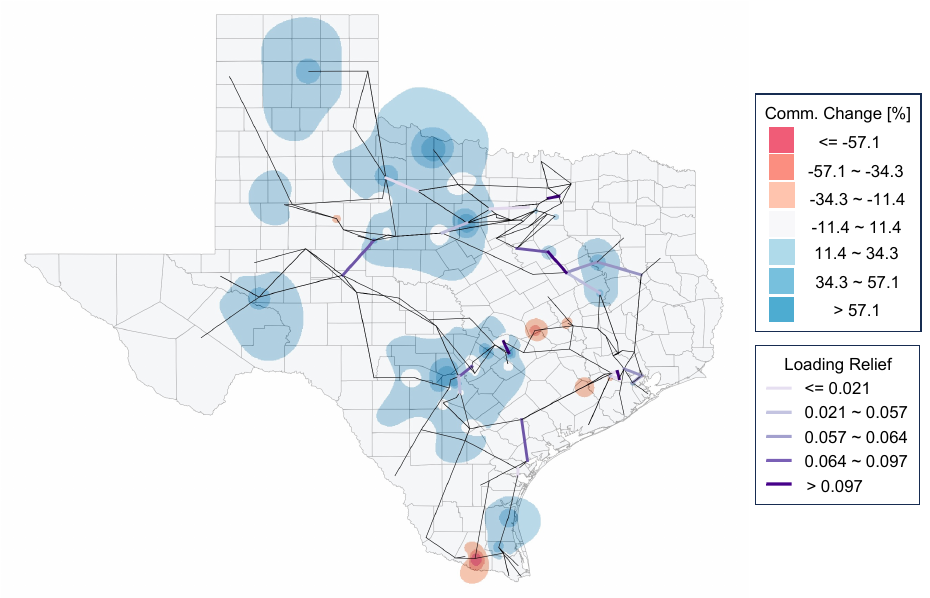}%
    \label{fig:QGIS_diff}}\\[1ex]
  \caption{\small (Case Study I) Annual heatmaps of day-ahead commitment and commitment changes on a map of Texas. (a) Annual day-ahead commitment from the DLR baseline, averaged over each bus; line colors indicate annual average loading (normalized by capacity).
  (b) Difference in commitment decisions in \%; blue indicates higher commitment under CHA-UC. Line colors show the corresponding reduction in loading (normalized by capacity) when applying CHA-UC instead of the DLR baseline.}
  \label{fig:QGIS}
  \vspace{-2mm}
\end{figure*}
\subsubsection{Benchmark Model}\label{subsec:QRF}

A quantile regression forest (QRF) model was employed to benchmark the performance of the CHA-UC model. QRF is an ensemble method that generates multiple decision trees from the training data and produces quantile forecasts from the forest output. The model setting was adopted from Dupin et al. \cite{Dupin2019}. The forest was created using 1-year data from 2022. The confidence level was set to 90\%, and the overestimation frequency was 7.4\%.
Overall, the quantile trajectories move with the observations, although sharp peaks are smoothed as a result of the ensemble-based quantile estimation. 
For the QRF-UC simulation, the DLR line constraints (\ref{Eq:DLR}) were replaced with QRF forecasts.

\subsection{Case study I - Value of Conductor Health Awareness} \label{sec:CS1}

In case study I, wind forecast errors are neglected ($\hat{P}^\text{wind}_{w,t} =P^\text{wind}_{w,t} $) and only DLR forecast errors are considered. 
This simplifies the risk structure and reduces the disparity between the day-ahead operation and the real-time operation, allowing a clearer interpretation of the results. 
Also, the minimum online up-reserve requirement is set to zero, as this case study does not account for WP forecast errors.

\subsubsection{Overall Performance Comparison}

Table~\ref{tab:exp_compare_grouped2} presents a breakdown of the normalized total cost of four models: baseline SLR, baseline DLR, QRF and CHA-UC. 
Day-ahead cost is the objective value of \eqref{Eq:baselineObj}, and total cost includes day-ahead, real-time, and conductor depreciation costs.
QRF-UC does not outperform baseline SLR: its lower day-ahead cost (21.29 vs. 21.44) is offset by higher depreciation cost (0.32 vs. 0.16); using a 95\% confidence level further increases total cost.
Although the DLR baseline yields the lowest day-ahead cost, followed by CHA-UC, QRF and SLR, CHA-UC achieves the lowest annual total cost of 21.44 M\$/day, reducing costs by 2.01\%, 0.75\% and 0.79\% compared to DLR, SLR and QRF, respectively. 
This improvement reflects its ability to balance the risk of ETO against reduced operation costs under DLR. 
Relative to SLR, WP curtailment is reduced by 85.8\% with DLR (11.7 vs. 82.6), 82.1\% with CHA-UC (14.8 vs. 82.6), and 67.9\% with QRF (26.5 vs. 82.6), showing that CHA-UC retains most of the congestion-relief of DLR.
On the other hand, depreciation costs were lowest for CHA-UC followed by SLR and DLR indicating that the CHA-UC preserves conductor lifespan most effectively. 
Finally, reserves are only activated in CHA-UC, as other benchmark baselines does not involve uncertainty in Case Study I. 

\subsubsection{Risk of ETO}

Figure \ref{fig:temp_year} visualizes the annual line temperature under the three models. In spring, fall and winter, the line temperatures were high in order of DLR baseline, CHA-UC and SLR baseline. 
This reflects the use of high ratings without stochastic concerns in DLR, high ratings with stochastic concerns in CHA-UC, and a uniformly low rating in SLR.
However, during summer, the overall line temperatures increase and the temperature of SLR exceeds CHA-UC. Since summer weathers are close to the conservative conditions for SLR calculation, the ``conservative'' SLR becomes risky as well. Thus, CHA-UC chooses a similarly conservative loading, shown in the day-ahead cost of 29.96 M\$/day for both CHA-UC and SLR Baseline.
A zoomed-in scatter plot confirms that SLR and DLR temperatures more frequently exceed the thermal limit than those of CHA-UC. This resulted in a CHA-UC depreciation cost of 0.12 M\$/day which was 3.8 times and 8.1 times less than SLR and DLR baseline, respectively. 

On the other hand, QRF chooses an even more conservative operation in summer, costing 30.35 M\$/day, reflected in Fig. \ref{fig:temp_year_qrf}. Yet this conservativeness was relatively ineffective, leading to a higher depreciation cost compared to CHA-UC (0.19 vs. 0.12). 
Despite the moving average of QRF and CHA-UC are similar, the QRF outliers show more frequent high temperature ETOs. The distribution in the zoom-in area shows that CHA-UC exploits the knowledge on the state of the conductor, balancing low-temperature ETOs with reduced operation costs, while QRF-UC frequently operates in higher temperature regions without conductor health-awareness. 
This is also reflected in the annual ETO statistics of all DLR-applied lines: the ETO frequency of QRF-UC is 1.2 times higher (2397 vs. 1934 hours) but the total depreciation cost is 2.6 times greater (0.32 vs. 0.12 M\$/day) due to the difference in average ETO temperatures (102.5 \textdegree C vs. 100.7 \textdegree C).
This result highlights the value of conductor-awareness of CHA-UC; conductor depreciation costs are minute albeit the smaller expenditure for conservativeness compared to the QRF benchmark model.

\subsubsection{Effect of CHA-UC to Commitment Decisions}

The annual average of commitment decisions made by DLR baseline is illustrated on a Texan map in Fig. \ref{fig:QGIS_u}. The cheapest units that comply with the UC and PF constraints are committed, leading to aggressive line loading. The employment of CHA-UC leads to committing additional costly units to hedge the risks of ETO, leading to an overall relief of line loading in DLR lines, as presented in Fig. \ref{fig:QGIS_diff}. For example, line 74, a key line connecting the south-west region and Dallas, is heavily loaded in a direction to Dallas (Bus 22 to Bus 94) with an annual average of 84.5\% (normalized by line capacity) when operated by DLR baseline. The transition to CHA-UC led to an 6.6\%p loading relief. This contributes to an annual conductor health relief of 13.6\%, which translates to a 108.8 M\$ reduction in depreciation costs. Here, the commitment decisions took a major role; the annual average commitment on bus 94 increased by 38\%, and the commitment on bus 30, adjacent to the east of bus 94, increased by 26.5\%, leading to the reduction of power flow to Dallas and stress on line 74.

\subsection{Case study II - Impact of Correlated Forecast Errors} \label{sec:CS2}
Case study I has shown the value of conductor health awareness. Case study II involves adding WP forecast errors to see the effects of correlated forecast errors. To handle wind generation forecast errors, a minimum reserve constraint of 5 GW (95\% VaR of 2022 WP forecast error) is enforced. 

\subsubsection{Overall Performance Comparison}

The costs of the annual simulation are reported on the right-hand side of Table \ref{tab:exp_compare_grouped2}. Overall, the trend of the three models from case study I was preserved. CHA-UC recorded the lowest annual total cost, 25.41 M\$, contributing to a 34.2\%, 24.7\% and 28.4\% reduction compared to DLR, SLR and QRF, respectively. However, unlike case study I, CHA-UC always incurred the highest day-ahead cost and it recorded the least reserve activation costs. This occurs because only CHA-UC accounts the stochasticity in WP forecasts in the day-ahead decision making.

\subsubsection{Impact of Correlated WP and DLR Forecast Error}

The effects from \textbf{Corollary} \ref{cor2} are studied and a physical interpretation is made.
Two pairs of wind farms and lines were selected for an analysis on inter-asset correlation. Table~\ref{tab:CS2_Pair} lists the rank correlation of pairs and average line flow using DLR-UC; each are a case of $s_l\mathrm{GSF}_{w,l}>0$ and $s_l\mathrm{GSF}_{w,l}<0$.

\begin{table}[t]
  \centering
    \captionsetup{justification=centering, labelsep=period, font=footnotesize, textfont=sc}
  \caption{ (Case Study II) Pairs of wind farms and lines. GSF was obtained by setting the paired bus as the from-bus.}
  \label{tab:CS2_Pair}
  \vspace{-2mm}
  \begin{tabular}{c|cc|ccc}
\toprule
Pair & Wind Farm & Line & Corr. & GSF & \makecell[l]{Avg. Lineflow [pu]}\\
\midrule
1 & 8  & 74  & 0.38 & 0.26 & -10.2\\
2 & 35 & 252 & 0.35 & 0.32 & 8.5\\
\bottomrule
\end{tabular}
\end{table}
\begin{figure}
  \centering
  \subfloat[Pair 1 (WP at Bus 22 and DLR on Line 74)]{%
    \includegraphics[width=\columnwidth]{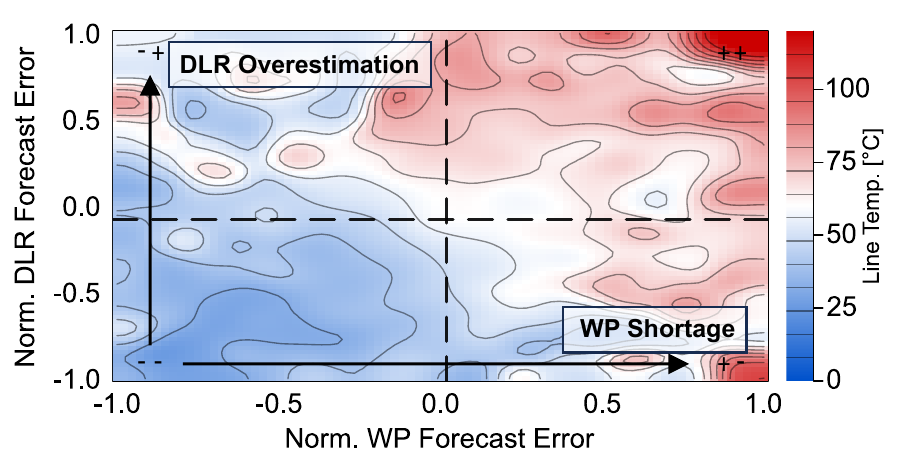}%
    \label{fig:heatmap1}%
  }\\
  \subfloat[Pair 2 (WP at Bus 120 and DLR on Line 252)]{%
    \includegraphics[width=\columnwidth]{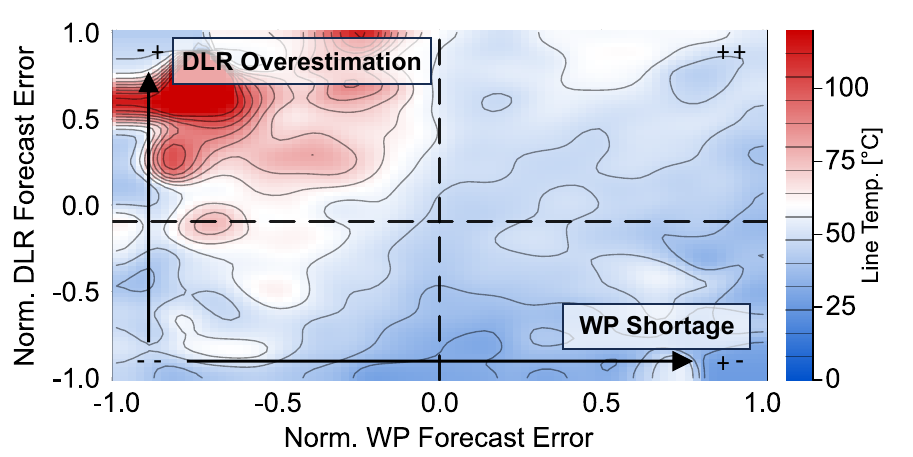}%
    \label{fig:heatmap2}%
  }
  \caption{\small (Case Study II) Heatmap of line temperatures under baseline DLR, given normalized WP and DLR forecast errors. Quadrants are indicated with pluses and minuses. a) Pair 1, b) Pair 2}
  \label{fig:heatmap}
  \vspace{-4mm}
\end{figure}
Figure \ref{fig:heatmap} shows the line-temperature heatmap under baseline DLR. Forecast errors are defined as forecast minus actual, so a larger error indicates greater overestimation. In pair 1, the positive GSF implies that a wind power (WP) shortage reduces line flow magnitude; because the line flow is negative, this means the flow moves away from zero, which stresses the conductor thermal limit. WP forecast errors are often mitigated by DLR forecast errors: DLR underestimation offsets WP shortage by providing additional line capacity in the fourth quadrant, whereas the first quadrant becomes risk-intensive because DLR overestimation amplifies the effect. In contrast, for pair 2, negative WP forecast errors (WP surplus) increase thermal stress, making the second quadrant risk-intensive. Positive correlation concentrates errors in specific quadrants, making pair 1 risk-prone and pair 2 risk-hedging, consistent with \textbf{Corollary}~\ref{cor2}.
\allowdisplaybreaks

\begin{table}[h]
  \centering
    \captionsetup{justification=centering, labelsep=period, font=footnotesize, textfont=sc}
  \caption{ (Case Study II) Annual average CHA-UC DA line flows with and without scenario correlation}
  \label{tab:CS2_CHAUC}
  \begin{tabular}{c|cc|c}
\toprule
Pair &
\makecell[l]{CHA-UC \\{} [pu]} &
\makecell[l]{CHA-UC w/o\\ Corr. [pu]} &
Conservativeness\\
\midrule
1 & -5.79 & -5.98 & Decreased\\
2 &  8.08 &  7.03 & Increased\\
\bottomrule
\end{tabular}
\end{table}

An experiment regarding correlation proves that the CHA-UC is successfully reflecting the emergent effect. The scenario indices of the pairs were scrambled, changing the marginal minimally, so that the correlation of the scenarios becomes close to 0. This will lead to the neglect of correlation. Separate experiments done with two pairs, in Table~\ref{tab:CS2_CHAUC}, show that the original (correlation-considering) CHA-UC has kept a more conservative line flow for the risk-prone pair 1 and a less conservative line flow for the risk-hedging pair 2. This asserts that the CHA-UC successfully handles the effect of correlation. 

\section{Conclusion}\label{sec:conclusion}

This paper presents a Conductor Health-Aware Unit Commitment (CHA-UC) framework that explicitly accounts for conductor degradation due to elevated temperature operation (ETO) under DLR. The proposed model integrates a tractable conductor temperature estimation module and a risk-based depreciation cost function, enabling the co-optimization of operating cost and long-term asset health.
Case studies demonstrate that CHA-UC reduces total cost by 0.75\% and wind power curtailment by 82\% compared to SLR operation. It also outperforms a baseline risk-averse approach (QRF-UC), while conventional DLR operation without risk consideration results in higher total costs due to excessive conductor degradation. 
This work also characterizes how the correlation between wind power and DLR forecast errors can amplify or hedge ETO-induced degradation based on grid conditions. This topology- and flow-dependent effect is explicitly captured by the CHA-UC model.
These findings motivate correlation-aware dispatch in operation and highlight the need to account for correlation structures when deploying DLR on wind-connected lines.

\section*{Acknowledgement}
The authors would like to thank Baptiste Rabecq and Andy Sun at the Massachusetts Institute of Technology for insightful discussions that sparked the initial idea for this work.

\appendix

\subsection{Proof of Proposition \ref{prop1}}\label{appendixA}
\begin{proof}
The variance of $\tilde{S}_{l,w}$ is
\begin{equation}
\mathrm{Var}(\tilde{S}_{l,w})
=
A_l^2\left(
\sigma_l^2+\mathrm{GSF}_{w,l}^2\sigma_w^2
-2s_l\mathrm{GSF}_{w,l}\rho_{w,l}\sigma_w\sigma_l
\right),
\end{equation}
where $\sigma_w$ and $\sigma_l$ are the standard deviations of $e_w$ and $e_l$, respectively. Differentiating with respect to $\rho_{w,l}$ gives
\begin{equation}
\frac{\partial \mathrm{Var}(\tilde{S}_{l,w})}{\partial \rho_{w,l}}
=
-2A_l^2 s_l\mathrm{GSF}_{w,l}\sigma_w\sigma_l. \label{Eq:VarDiff}
\end{equation}
Since $A_l^2$, $\sigma_w$, and $\sigma_l$ are positive, \eqref{Eq:VarDiff} is positive when $\mathrm{GSF}_{w,l} p_l < 0$, vice versa.
Given $s_l=\mathrm{sign}(p_l)$, the result is equivalently stated by the sign of $\mathrm{GSF}_{w,l}p_l$.
\end{proof}

\subsection{Proof of Corollary \ref{cor2}}\label{appendixB}

\begin{proof}
The conductor depreciation cost function $C_l^\mathrm{Dep}(\tau_l)$ is convex and non-decreasing, and $\tau_l$ is affine in $\tilde{S}_{l,w}$, 
Under fixed marginal distributions, changing $\rho_{w,l}$ changes the dispersion of $\tilde{S}_{l,w}$ without changing its mean. By Proposition~\ref{prop1}, this dispersion increases when $\mathrm{GSF}_{w,l}p_l<0$ and decreases when $\mathrm{GSF}_{w,l}p_l>0$. Since $C_l^\mathrm{Dep}(\tau_l)$ is non-decreasing and convex, the result follows directly from lemma \ref{lem}.
\end{proof}

\bibliographystyle{IEEEtran}
\bibliography{chauc.bib}

\end{document}